# Towards Fast Evaluation of Unsupervised Link Prediction by Random Sampling Unobserved Links

Jingwei Wang [a], Yunlong Ma [a,*], Yun Yuan [a]

[a] College of Electronics and Information Engineering, Tongji University, Shanghai, China.
* Correspondence to: evanma@tongji.edu.cn.

## Abstract

Link prediction has aroused extensive attention since it can both discover hidden connections and predict future links in the networks. Many unsupervised link prediction algorithms have been proposed to find these links in a variety of networks. However, there is an evaluation conundrum in unsupervised link prediction. Unobserved links heavily outnumber observed links in large networks, so it is unrealistic to quantify the existence likelihood of all unobserved links to evaluate these algorithms. In this paper, we propose a new evaluation paradigm that is sampling unobserved links to address this problem. First, we demonstrate that the proposed paradigm is feasible in theory. Then, we perform extensive evaluation experiments in real-world networks of different contexts and sizes. The results show that the performance of similarity-based link prediction algorithms is highly stable even at a low sampling ratio in large networks, and the evaluation time degradation caused by sampling is striking. Our findings have broad implications for link prediction.

## 1. Introduction

Link prediction is a basic problem in network science, aiming to estimate the connection likelihood of unconnected node pairs based on the observed network [1]. It has recently aroused broad attention. Since researchers have found that it can model many real-world problems and discover hidden connections [2, 3] or predict new links [4, 5], such as traffic flow prediction [6], polypharmacy side effects discovery [7], drug combinations discovery [8], relation completion in knowledge base [9], item recommendation [10] and so forth. So far, various link prediction methods have been put forward [4], including similarity-based methods [2] and supervised learning-based methods [11].

Here, we focus on the evaluation of link prediction methods. According to the evaluation paradigms, link prediction can be divided into two categories: unsupervised link prediction [12] and supervised link prediction [13]. In unsupervised link prediction, to evaluate the performance of an algorithm, the possibilities of the existence of all unobserved links in a network need to be calculated, but it is infeasible for large networks. Because an undirected network with $n$ nodes and $m$ observed links contains $n(n-1)/2-m$ unobserved links which heavily outnumber observed links. Taking the famous Karate network as an example, Figure 1 shows the observed links and the unobserved links in this network. The observed links are rare and sparse, whereas the unobserved links are the overwhelming majority. Furthermore, Figure 2 shows how the number of unobserved links and the evaluation time soar with increasing network size. The number of unobserved links grows quadratically with the increase of the number of nodes. It indicates the computational complexity for the evaluation of unsupervised link prediction is $O(n^2)$. This is in line with the actual evaluation time. In addition to time, massive unobserved links in large networks also occupy a lot of storage space, which is beyond the capacity of a single computer. This conundrum seriously limits the evaluation of unsupervised link prediction algorithms in large networks.

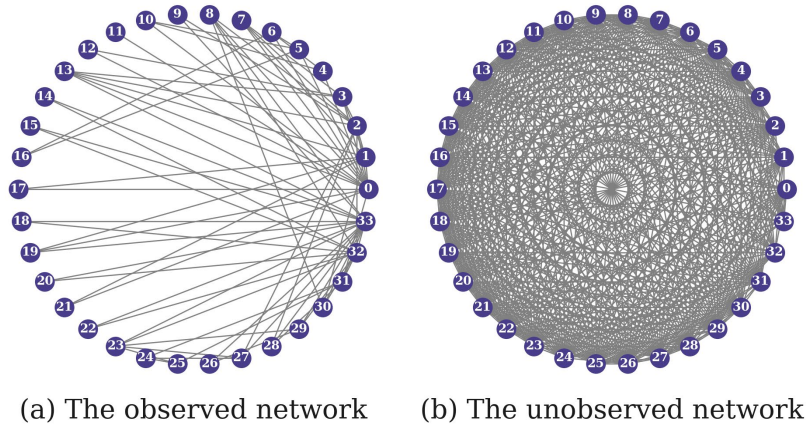

(a) The observed network    (b) The unobserved network

Figure 1. Karate network. (a) The observed network consists of 34 nodes and 78 observed links. (b) The unobserved network consists of 34 nodes and 483 unobserved links.

Some recent studies have also pointed out this problem [14, 15, 16, 17], but their attention is either to find faster algorithms or to reduce the size of observed networks. The former is still unable to avoid this problem, while the latter is at the cost of loss of network structural information. Besides, some researchers tend to evaluate link prediction methods in a supervised way [13, 18, 19]. In supervised link prediction, the observed links and the unobserved links are considered as positive instances and negative instances, respectively. Link prediction thereby is regarded as a binary classification task with unbalanced categories. Hence, negative sampling is introduced to solve this class imbalance problem [20, 21]. This supervised evaluation is inapplicable to non-parametric unsupervised link prediction methods, such as similarity-based ones. However, if it is feasible to sample unobserved links in unsupervised link prediction, the evaluation conundrum mentioned above will be addressed.

Previous studies pay most attention to the effect of negative sampling on the evaluation of supervised link prediction. It is unclear whether sampling unobserved links can be used in the evaluation of unsupervised link prediction. Here, we address this question through theoretical analysis and empirical analysis. The main contributions of this paper can be summarized as follows.

(1) To address the evaluation conundrum of unsupervised link prediction, we propose a new evaluation method, which can reduce the evaluation time and ensure the effectiveness of evaluation by sampling unobserved links.
(2) We demonstrate the sampling ratio can decrease as the network size grows, and the effect of sampling unobserved links on the evaluation of unsupervised link prediction becomes very small in large networks in theory.
(3) We perform extensive evaluation experiments on real-world networks with various unsupervised link prediction algorithms. Our results indicate that the performance of these algorithms shows high stability for sampling unobserved links, especially in large networks, and the rankings of these algorithms are nearly insensitive to the sampling ratio. Moreover, the evaluation time of algorithms reduces remarkably as the sampling ratio declines. We can strike the right balance between performance and time.

The rest of the paper is organized as follows. Section 2 reviews some related works, including unsupervised link prediction and supervised link prediction. Link prediction problem, various unsupervised link prediction algorithms, and the evaluation metric are illustrated in Section 3. Section 4 presents the theoretical analysis and complexity analysis of sampling unobserved links, as well as the sampling method. The experimental results and the detailed analysis are reported in Section 5, and Section 6 gives the conclusion.

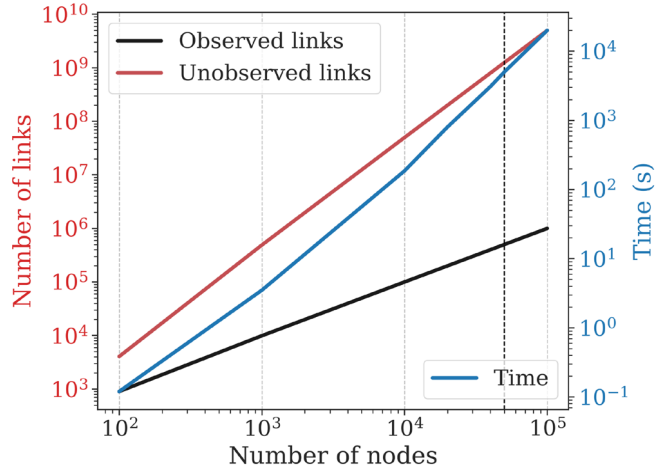

Figure 2. Evaluation time of preferential attachment (PA) algorithm in Barabási–Albert (BA) networks with different sizes. Black line: the number of observed links. Red line: the number of unobserved links. Blue line: the evaluation time of PA in unobserved links. Black dashed line: the system is out of memory. The evaluation time is predicted when the number of nodes exceeds 40000.

## 2. Related work

In the last decade, a great number of link prediction methods have been proposed. Most of these methods are unsupervised but recently supervised learning-based methods have attracted increasing interest. This section briefly reviews some important work and new developments of unsupervised link prediction and supervised link prediction, as well as some methods based on the big data platforms.

*2.1 Unsupervised link prediction*

There are two evaluation paradigms in unsupervised link prediction. The first paradigm is evaluation in temporal networks proposed by David Liben-Nowell and Jon Kleinberg [22]. They first define the link prediction problem and evaluate link prediction methods in temporal coauthorship networks. In particular, a temporal network is divided into two subnetworks according to the time sequence. Then, the subnetwork of the previous period is used as the training set to predict which links will appear in the subnetwork of the next period. This paradigm is popular in temporal domains, such as scientific collaboration [2].

Since the networks having such incremental nature are rare and many network datasets without timestamps, most studies adopt the other paradigm that is evaluation in static networks. Yet, we do not know which links are future links in static networks. To test an algorithm's accuracy, the observed links are randomly divided into two parts: the training set is treated as known information, while the probe set (i.e., validation subset) is used for testing and no information in this set is allowed to be used for prediction. This evaluation paradigm is proposed by Linyuan Lü et al. [1, 23] and has become the first choice.

However, no matter what kind of evaluation paradigm, this evaluation conundrum exists. To solve this problem, researchers have proposed some sampling methods to reduce the number of links to be predicted. The existing sampling methods can be classified into two groups. The first group aims at sampling the observed links to obtain a representative subnetwork of the original network with fewer nodes. The random walk is a commonly used sampling method. In [23] Lü et al used random walk sampling to obtain three subnetworks (4000 nodes) to represent the original Arxiv network (22908 nodes), Facebook network (56952 nodes), and Enron network (87273 nodes) respectively. Ahmed et al. [24] firstly constructed a subgraph centered at each node in the weighted graph by a random walk from the node and then calculated the similarity score within such small sub-graphs. However, it is found in [25, 26] that the topological properties in sampled networks might be estimated quite differently for different sampling methods.

The second group focuses on some links related to nodes of interest or some disconnected node pairs at one certain distance. Liben-Nowell and Kleinberg [22] investigated the prediction of new collaborations at distance 2 or 3 in five co-authorship networks. Similar to [22], these works [27, 28] focus on the prediction of pairs of nodes that are 2-hop or 3-hop away in the network, i.e., the shortest path length between two nodes is 2 or 3. Chen et al. [16] proposed a fast similarity-based method to predict links related to nodes of interest; i.e., they just considered those links that users are interested in rather than all node pairs. Duan et al. [14] proposed an ensemble-enabled approach for top-$k$ link prediction, which scales up link prediction on very large social networks. But the disadvantage of this algorithm is that there are too many parameters to be optimized manually.

*2.2 Supervised link prediction*

Mohammad Al Hasan et al. [18] first study link prediction as a supervised learning task, namely a binary classification task mentioned in the Introduction. This work provides an excellent starting point for supervised link prediction. Recently, many studies from the machine learning community [6, 7, 11, 19, 29, 30] prefer supervised learning for link prediction. In these works, negative sampling is used to obtain negative instances, and the number of negative instances should be as many as that of positive instances to keep class balance. For example, if 10% of observed links are removed randomly from the network as positive testing links, the same number of unobserved links will be sampled randomly as negative testing links; then, the remaining 90% of observed links as well as the same number of other unobserved links to construct the train set. Inspired by negative sampling in supervised link prediction, we introduce sampling unobserved links into the evaluation of unsupervised link prediction.

In short, the supervised evaluation is appropriate to parametric link prediction methods such as graph neural network-based algorithms [11, 31, 32], because the parameter can be adjusted by gradient descent and other optimization methods. Besides, it is convenient to incorporate topological features and non-topological features to perform link prediction in supervised link prediction.

*2.3 Big data platform*

For large-scale networks, it is a natural idea to use big data platforms, such as MapReduce, Spark, and Flink, to cope with this problem. These platforms generally provide a framework to deal with graph data, such as PEGASUS in MapReduce, GraphX in Spark, Gelly in Flink. Cui et al. [15] proposed a fast algorithm based on MapReduce to obtain the common neighbors (CN) number of all node pairs. Yang et al. [33] proposed an algorithm based on the local neighbor link (LNL) for large-scale networks and implemented it in both MapReduce and Spark. The results show that implementation by Spark has higher efficiency than using MapReduce. Katragadda et al. [34] studied the distributed link prediction problem in dynamic graph streams. They propose a neighborhood-centric graph processing approach to handle this problem that exploits the locality, parallelism, and incremental computation of a distributed framework, namely Flink.

However, the cost of building or renting a distributed computing platform is extremely expensive, which is unaffordable for many research groups. Moreover, these platforms generally adopt distributed computing to achieve parallel link prediction, bringing a problem that it is difficult to rank testing links globally, which means that it is impossible to evaluate a link prediction algorithm with the precision metric [35].

## 3. Preliminaries

*3.1 Problem description*

Consider an undirected network $G(V, E)$, where $V$ is the set of nodes and $E$ is the set of observed links (or edges) among the nodes. Multiple links and self-loop are not allowed. For each node $x$, the number of edges connected to it is defined as its degree $k_x$; then the averaged degree of $G$ is $\langle k \rangle = 2|E|/|V|$, where $|E|$ and $|V|$ are the number of the links and the nodes, respectively. $\langle d \rangle$ is the average shortest distance between node pairs in $G$. $C$ is the cluster coefficient of $G$ [36]. The set of common neighbors of nodes $x$ and $y$ is denoted by $\Lambda_{x,y} = \Gamma(x) \cap \Gamma(y)$, where $\Gamma(x)$ denotes the set of neighbors of node $x$.

The universal set denotes by $U$, containing all $\frac{|V|(|V|-1)}{2}$ possible links in $G$. Then, the set of unobserved links is $N = U - E$. According to the paradigm of link prediction [1], the observed links, $E$, are randomly divided into two parts: the training set, $E^T$, and the probe set, $E^P$. It is clear that $E^T \cup E^P = E$ and $E^T \cap E^P = \emptyset$. $E^P$ and $N$ together constitute the testing set, denoted by $T$. In unsupervised link prediction, one must evaluate the existence likelihood of all links in $T$. Here, each link $(x, y) \in T$, where $x, y$ are a pair of disconnected nodes, will be assigned a score $S_{x,y}$ by a algorithm. The likelihood of connecting nodes $x, y$ is increasing as the increase of the score, and vice versa. The goal of link prediction is to make the score of the probe links in $E^P$ higher than that of the unobserved links in $N$ as much as possible. **Table 1** summarizes the concepts and its corresponding mathematical symbols in this paper.

**Table 1**
Concepts and mathematical symbols.

| Symbol | Concept |
|---|---|
| $G$ | An undirected network |
| $V$ | The set of nodes in $G$ |
| $E$ | The set of observed links in $G$ |
| $k_x$ | The degree of node $x$ |
| $\langle k \rangle$ | The averaged degree of $G$ |
| $\langle d \rangle$ | The average shortest distance of $G$ |
| $C$ | The cluster coefficient of $G$ |
| $\Gamma(x)$ | The set of neighbors of node $x$ |
| $x, y$ | A pair of nodes |
| $\Lambda_{x,y}$ | The set of common neighbors of $x, y$ |
| $U$ | The set of all possible links in $G$ |
| $N$ | The set of unobserved links in $G$ |
| $E^T$ | The training set |
| $E^P$ | The probe set |
| $T$ | The set of testing links |
| $(x, y)$ | A link |
| $S_{x,y}$ | The existence likelihood of link $(x, y)$ |
| $s$ | The sampling ratio of $N$ |
| $N_s$ | The set of links sampled from $N$ |
| $|\cdot|$ | The number of elements of the set |

*3.2 Link prediction algorithms*

Many algorithms have been proposed for unsupervised link prediction. In this study, we evaluate seven well-known similarity based algorithms, because they have high accuracy and low computational complexity.

(1) Common Neighbors (CN) algorithm [37]. It is assumed that two nodes with more common neighbors are easily connected. Then the score of an edge could be defined as

$$S_{x,y}^{CN} = |\Gamma(x) \cap \Gamma(y)|.$$

(2) Jaccard (Jac) algorithm. The score of each pair could also be obtained from Jaccard's definition as

$$S_{x,y}^{Jac} = \frac{|\Gamma(x) \cap \Gamma(y)|}{|\Gamma(x) \cup \Gamma(y)|}.$$

(3) Leicht-Holme-Newman algorithm (LHN) [38]. This index assigns high similarity to node pairs that have many common neighbors compared not to the possible maximum, but to the expected number of such neighbors. It is defined as

$$S_{x,y}^{LHN1} = \frac{|\Gamma(x) \cap \Gamma(y)|}{k_x \times k_y}.$$

(4) Adamic-Adar (AA) algorithm [39]. This index refines the simple counting of common neighbors by assigning the less-connected neighbors more weights, and is defined as

$$S_{x,y}^{AA} = \sum_{z \in \Gamma(x) \cap \Gamma(y)} \frac{1}{\log k_z}.$$

(5) Resource Allocation (RA) algorithm [40]. This index is motivated by the resource allocation dynamics on complex networks. The similarity between and can be defined as the amount of resource $y$ received from $x$, which is

$$S_{x,y}^{RA} = \sum_{z \in \Gamma(x) \cap \Gamma(y)} \frac{1}{k_z}.$$

(6) Preferential Attachment (PA) algorithm [41]. The mechanism of preferential attachment can be used to generate evolving scale-free networks, where the probability that a new link is connected to the node $x$ is proportional to $k_x$. It is defined as

$$S_{x,y}^{PA} = k_x \times k_y.$$

The above six algorithms are based on local structure similarity, namely neighborhood information. So they are also called local algorithms. Besides, we consider one more complex *quasi-local* algorithm, which do not require global topological information but make use of more information than *local* algorithms. In other words, it provides a good tradeoff of accuracy and computational complexity.

(7) The Local Path (LP) algorithm [42]. LP takes consideration of 2-order paths and 3-order paths, and is defined as

$$S_{x,y}^{LP} = A^2 + \sigma A^3.$$

where $\sigma$ is a free parameter. Clearly, $(A^2)_{xy}$ is equal to the CN predictor and $(A^3)_{xy}$ denotes the number of paths of length 3 connecting $x$ and $y$. Here, the parameter $\sigma$ is set as 0.01 according to [1], instead of finding out its optimum that may cost much time.

*3.3 Evaluation metric*

Two standard metrics are widely used to quantify the accuracy of algorithms in unsupervised link prediction: area under the receiver operating characteristic curve (AUC) and Precision. The AUC evaluates the algorithm's performance according to the whole list of all unobserved links while the precision only focuses on the $L$ links with top ranks or highest scores. Since the precision is a fixed-threshold metric, it will be affected by the number change of unobserved links. So it is inapplicable to the evaluation with sampling, this has been pointed out in previous studies [43, 44]. We only consider the AUC in this paper. Note, the calculation of AUC is different from that of AUC in the machine learning community. It is defined by Linyuan Lü and Tao Zhou in [1]. A detailed introduction of the AUC is as follows.

Given the score of all testing links, the AUC can be viewed as the probability that a randomly chosen probe link (i.e., a link in $E^P$) is given a higher score than a randomly chosen unobserved link (i.e., a link in $E^S$). In the algorithmic implementation, we usually calculate the score of each non-observed link instead of giving the ordered list since the latter task is more time-consuming. Then, at each time we randomly pick a missing link and a nonexistent link to compare their scores, if among $n$ times independent experiments, there are $n_1$ times when the probe link has a higher score than the non-observed link and $n_2$ times when they have the same score, the AUC value is

$$AUC = \frac{n_1 + 0.5 n_2}{n}.$$

If all the scores are generated from an independent and identical distribution, the accuracy should be about 0.5. Therefore, the degree to which the accuracy exceeds 0.5 indicates how much better the algorithm performs than pure chance. In this paper, we set $n = 100000$.

## 4. Methodology

In this section, we first introduce the proposed evaluation paradigm for unsupervised link prediction, and then provide the mathematical proof for the feasibility of sampling unobserved links in the evaluation. Finally, we analysis the computational complexity of this new evaluation paradigm.

*4.1 Evaluation paradigm*

In the traditional evaluation paradigm, all unobserved links are treated as testing links. This is the crux of the evaluation conundrum. Therefore, we propose a new evaluation paradigm that only samples a part of unobserved links as testing links. Figure 3 shows the traditional evaluation paradigm and proposed evaluation paradigm. The difference between the two paradigms lies in the number of unobserved links included in the testing links.

Since the structure of a network is dependent on the observed links rather than the unobserved links. Sampling the unobserved links does not affect the network structure, so it will not cause the loss of structure information. That is its advantage over the sampling subnetwork method. As mentioned in Section 2.1, Lü et al [23] used random walk sampling to get a subnetwork to apply their proposed SPM to large networks. However, the number of nodes in a subnetwork is less than 20% of that in the original network, leading to a serious loss of structure information. Thus, the sampled subnetwork cannot represent the original network.

Here, we use pure random sampling to obtain samples from the unobserved set as testing links. Because only pure random sampling can lead to unbiased evaluation. The set of sampled unobserved links denotes by $N_s$. In particular, given a sampling ratio $s$, we randomly select $s|N|$ unique links from $N$ and add them to $N_s$. The number of unobserved links sampled reduces with the decrease of the sampling ratio. For any network, the minimum sampling ratio should ensure that the number of unobserved links sampled is greater than 0.

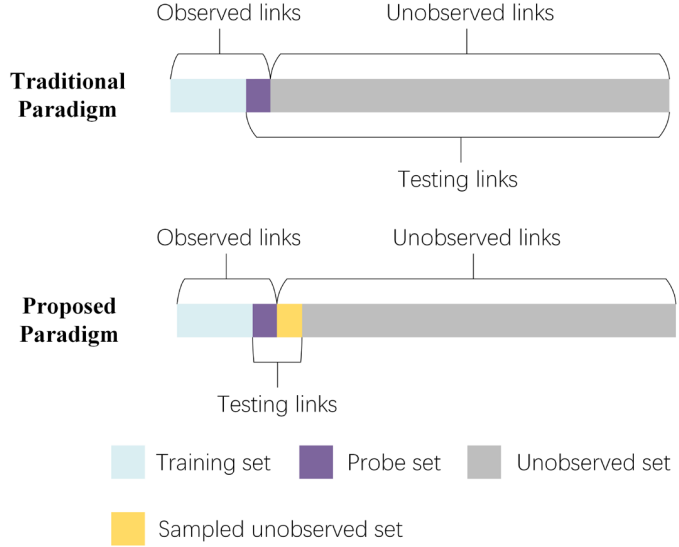

Figure 3. Traditional evaluation paradigm and the proposed evaluation paradigm.

The general framework of proposed evaluation paradigm is presented in Algorithm 1, which mainly consists of the following four steps. First, the observed links of the input network are divided into the train set $E^T$ and the probe set $E^P$. Here, we ensure the train network generated by $E^T$ is connected, because similarity-based methods have requirements for connectedness. Second, we random sample unobserved links from $N$ at the sampling ratio $s$ to obtain $N_s$. The testing links consist of $E^P$ and $N_s$. Third, the similarity score of links in $E^P$ and $N_s$ is assigned by each link prediction algorithm based on the train graph, and then the AUC of each algorithm is calculated. Finally, the AUC values of different algorithms are compared to evaluate the performance of these algorithms.

**Algorithm 1** Proposed evaluation paradigm for unsupervised link prediction

**Input:** $G(V, E)$: a complex network;
$s$: the sampling ratio of unobserved links;
$\mathcal{L}_1, \dots, \mathcal{L}_p$: a series of link prediction algorithms.
**Output:** $AUC_1, \dots, AUC_p$: the AUC of link prediction algorithms $\mathcal{L}_1, \dots, \mathcal{L}_p$.
1: $N \leftarrow$ the unobserved links in $G$
2: $E^T, E^P \leftarrow$ split observed links in $G$ into train set and probe set
3: $G_{train} \leftarrow$ generate a connected train network by $E^T$
4: $N_s \leftarrow$ sample unobserved links from $N$ at the sampling ratio $s$
5: **for** $i$ **in** $\mathcal{L}_1, \dots, \mathcal{L}_p$:
6:   $Sim(E^P) \leftarrow$ get the similarity score of links in $E^P$ by $\mathcal{L}_i$ based on $G_{train}$
7:   $Sim(N_s) \leftarrow$ get the similarity score of links in $N_s$ by $\mathcal{L}_i$ based on $G_{train}$
8:   $AUC_i \leftarrow$ calculate the AUC of $\mathcal{L}_i$ based on $Sim(E^P)$ and $Sim(N_s)$
9: **end for**
10: **return** $AUC_1, \dots, AUC_p$

*4.2 Theoretical analysis*

To prove that sampling unobserved links is feasible in the evaluation of unsupervised link prediction, we provide a concrete mathematical proof as bellows.

**Theorem 1**. *For any link prediction algorithm $\mathcal{L}$, the performance variance of $\mathcal{L}$ increases when the sampling ratio $s$ of unobserved links decreases.*

*Proof.* For a specific link prediction algorithm $\mathcal{L}$, we assume that among all $|N|$ unobserved links there are $M$ links that can be predicted correctly by $\mathcal{L}$ while the other $|N| - M$ links cannot be predicted correctly by $\mathcal{L}$.

Based on the fact that we randomly sample $|N_s|$ unobserved links for inclusion in the final test set, the number of unobserved links that can be predicted correctly by $\mathcal{L}$ among these $|N_s|$ unobserved links is a random variable X that has probability mass function:

$$P(X = x) = \frac{\binom{M}{x}\binom{|N|-M}{|N_s|-x}}{\binom{|N|}{|N_s|}}.$$

Obviously, $X$ follows a hypergeometric distribution. The expectation and the variance of $X$ are:

$$E(X) = \frac{|N_s|M}{|N|},$$

$$Var(X) = \frac{M(|N|-M)s(1-s)}{(|N|-1)}.$$

Hence, $\frac{X}{|N_s|}$ has expectation and variance following:

$$E\left(\frac{X}{|N_s|}\right) = \frac{M}{|N|}, \tag{1}$$

$$Var\left(\frac{X}{|N_s|}\right) = \frac{M(|N|-M)(1-s)}{|N|^2(|N|-1)s}. \tag{2}$$

Clearly, the performance expectation of $\frac{X}{|N_s|}$ will not change with the change of sampling ratio, but the performance variance of $\frac{X}{|N_s|}$ will increase with the decrease of sampling ratio. The proof is complete.

**Theorem 2**. *For a given sampling ratio $s$, the performance variance of a link prediction algorithm $\mathcal{L}$ decreases when the number of unobserved links increases.*

*Proof.* Based on Theorem 1, let $H = \frac{M(|N|-M)}{|N|^2(|N|-1)}, M \in [0, |N|]$. It is easy to obtain the upper bound and lower bound of $H$, that is $0 \leq H \leq \frac{1}{4(|N|-1)}$.

Since the sampling ratio $s$ is a constant, $Var\left(\frac{X}{|N_s|}\right)$ is only dependent on the number of unobserved links $|N|$, and the maximum of $Var\left(\frac{X}{|N_s|}\right)$ is $\frac{1-s}{4(|N|-1)s}$. Consequently, $Var\left(\frac{X}{|N_s|}\right)$ decreases when $|N|$ increases.

**Theorem 3**. *For a given variance $\mathcal{V}$, the sampling ratio $s$ of unobserved links can reduce as the number of unobserved links increases.*

*Proof.* Based on Theorem 1 and Theorem 2, suppose $Var\left(\frac{X}{|N_s|}\right) = \frac{M(|N|-M)(1-s)}{|N|^2(|N|-1)s} < \mathcal{V}$, that is $H \cdot \frac{1-s}{s} \leq \mathcal{V}$. Then, one has $s \geq \frac{H}{H+\mathcal{V}}$. Since $\mathcal{V}$ is a constant ($\mathcal{V} > 0$), $\frac{H}{H+\mathcal{V}}$ has the maximum when $H = \frac{1}{4(|N|-1)}$. Thus, one has

$$s \geq \frac{1}{4\mathcal{V}(|N|-1)+1}, \tag{3}$$

it follows that the minimum of $s$ decreases when $|N|$ increases. Completing this proof.

Furthermore, since the number of unobserved links increases with the growth of number of nodes in networks, it is easy to obtain the following two corollaries.

**Corollary 1**. *Given a constant sampling ratio, the performance variance of a link prediction algorithm decreases when the number of nodes in networks increases.*

**Corollary 2**. *Given a constant variance, the sampling ratio of unobserved links can reduce when the number of nodes in networks increases.*

Taken together, we demonstrate the sampling ratio can decrease as the network size grows, and the effect of sampling unobserved links on the evaluation of unsupervised link prediction becomes small in large networks in theory. This establishes a concrete theoretical foundation for our following experiments.

*4.3 Complexity analysis*

The main purpose of this work is to address the evaluation conundrum in unsupervised link prediction, that is to reduce the time complexity of evaluation. Since the time complexity of the link prediction algorithm has been discussed in [1], it is not the focus of this paper.

As mentioned in the Introduction, the time complexity of the traditional evaluation paradigm is $O(|V|^2)$. According to our theoretical analysis, we can adopt a low sampling ratio in large networks, so the time complexity of the proposed evaluation paradigm will far below that of the traditional paradigm. As we show next, when the number of unobserved links sampled decreases to the number of probe links in large networks, the evaluated performance of an algorithm remains stable. Our empirical results show that the time complexity of our proposed evaluation paradigm can be as low as $O(|E|)$ and even $O(|V|)$. That is a remarkable improvement. Consequently, the proposed evaluation paradigm helps unsupervised link prediction to escape the evaluation dilemma and enables many new proposed unsupervised methods to be quickly verified in large networks.

## 5. Experimental Evaluation

*5.1 Datasets and experimental settings*

In this section, eight real-world networks used are depicted in detail. According to the setting in [1], we conduct experiments only on the giant connected components of these networks, and all of them are treated as undirected and unweighted.

The eight networks come from disparate fields. Karate is a famous social network of friendships from a karate club in Zachary [36]. Dolphin is a network of frequent associations between dolphins [45]. Word is a network of word adjacencies of common adjectives and nouns in the novel "David Copperfield" by Charles Dickens [46]. Metabolic is a network of chemical reactions in Celegans, where nodes represent the chemicals and edges are reactions among them [47]. Email is a network representing the exchange of emails among members of a university in Spain in 2003 [48]. TVShows and Company are Facebook page networks of different categories [49], where nodes represent the pages and edges are mutual likes among them. LastFM is a social network of LastFM users that was collected from the public API in March 2020 [50]. Nodes represent LastFM users from Asian countries and edges represent mutual follower relationships between them. The topological information of these networks is described in **Table 2**. |V| represents the number of nodes, |E| represents the number of edges, and |N| represents the number of unobserved edges. <k>, C, and <d> represent the average degree, the clustering coefficient, and the diameter of a network, respectively. It is noteworthy that the number of nodes in these networks ranges from 34 to 14113. Such a wide range enables us to see how the sampling ratio interacts with the network size.

**Table 2**

Topological information of eight real-world networks.

| Networks | |V| | |E| | |N| | <k> | C | <d> |
|---|---|---|---|---|---|---|
| Karate | 34 | 78 | 483 | 4.59 | 0.57 | 2.41 |
| Dolphin | 62 | 159 | 1732 | 5.13 | 0.26 | 3.36 |
| Word | 112 | 425 | 5791 | 7.59 | 0.17 | 2.54 |
| Metabolic | 453 | 2025 | 100353 | 9.01 | 0.65 | 2.66 |
| Email | 1133 | 5451 | 635827 | 9.62 | 0.22 | 3.61 |
| TVShows | 3892 | 17239 | 7554647 | 8.87 | 0.37 | 6.26 |
| LastFM | 7624 | 27806 | 29031070 | 7.29 | 0.22 | 5.23 |
| Company | 14113 | 52126 | 99529202 | 7.41 | 0.24 | 5.31 |

For each network, we random sample unobserved links for the evaluation of unsupervised link prediction. Since the number of unobserved links varies greatly in networks of different sizes (see Table 1), we set different sampling ratios for different networks. In all networks, the maximum sampling ratio is 1, and the minimum sampling ratio can ensure the number of sample links not less than 1. The sampling ratio is the proportion of unobserved links sampled to all unobserved links.

In our experiments, we keep $|E^T|:|E^P| = 9:1$ for all networks. As mentioned in Section 4.2, some methods have requirements such as connectedness of the train network, it is unfeasible to random split the train set and the probe set. Here, we adopt a naïve approach to split the input network and derive a connected train network in which the edges can span all nodes in the input network. In particular, we take the input network as the train network at first. Then, we randomly remove a link from the train network and see if the remaining train network is connected. If yes, we add this link to the probe set; otherwise, the link is added to the train set. We repeat this process until the number of links in the probe set meets the requirements. Results in each network are obtained by averaging over 100 implementations with this division approach. All experiments are performed with Python3 on a computer with an Intel Xeon 48-Core 2.5GHz CPU, 256 GB RAM, and Ubuntu 18.04.2 system.

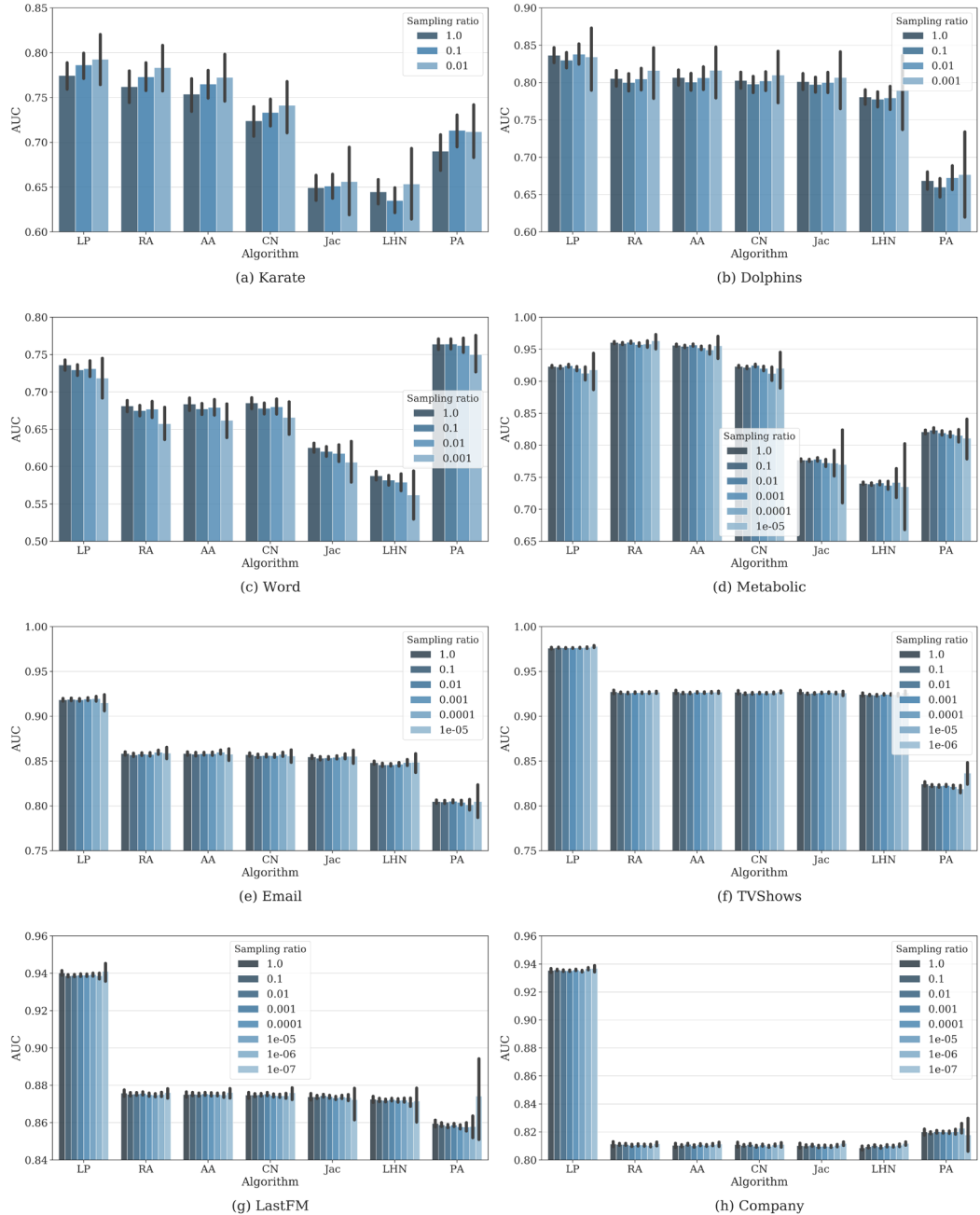

Figure 4. AUC of seven algorithms in eight networks under different ratios. The height of each rectangle represents the mean of AUC, and the deep line on each rectangle represents the error bar.

*5.2 AUC*

First, we focus on the effect of different sampling ratios on the performance of link prediction algorithms in networks. Figure 4 shows the AUC of seven algorithms in eight networks under different ratios. There are two striking results in Figure 4. First, in small networks (Karate, Dolphin, Word, and Metabolic), the AUC of link prediction algorithms becomes shaky when the sampling ratio declines; while in large networks (Email, TVShows, LastFM, and Company), the performance is insensitive to the change of the sampling ratio. That shows that the increase of network size gradually weakens the influence of sampling unobserved links, which is in line with our theoretical demonstration (see Theorem 3 and Corollary 2). We will further discuss the relationship between network size and sampling ratio in the next two subsections.

Second, the AUC of the neighborhood-based algorithms (CN, Jac, LHN, AA, and RA) is distinct in small networks but very similar in large networks. That occurs because of the following two reasons. The first is that neighborhood-based algorithms have the problem of "degeneracy of the states" proposed in [1]. These algorithms only use the common neighbor information that is less distinguishable, so they have similar results in predicting the scores of probe links. The other reason is that the average shortest distances of the four large networks are greater than that of the other networks, which are more than 3 (see Table 2). As a result, most unobserved links are assigned zero by these neighborhood-based algorithms because there is no common neighbor between two nodes of an unobserved link whose shortest distance is more than 3. Therefore, the AUC of these algorithms is mainly dependent on the scores of probe links. Taken together, we can explain why the neighborhood-based algorithms have very similar performance in large networks.

Besides, it has been found that using a little bit more information involving the next nearest neighbors can break the "degeneracy of the states" and make the similarity scores more distinguishable. That is the reason why LP improves the AUC and performs best in large networks. As for PA, it has the worst overall performance, since it requires the least information. The only exception is that it performs best in the Word network. In fact, in natural language, high-frequency words are always used together with high-frequency words. In network science terms, high-degree nodes tend to link to other high-degree nodes. That is the prediction mechanism of PA so that it outperforms others in Word.

*5.3 Algorithm ranking*

Although the AUC of link prediction algorithms is affected by the sampling ratio in small networks, we find another stability of these algorithms. Figure 5 shows the performance change of the seven algorithms in four small networks, and we can observe the rankings of these algorithms under different sampling ratios. Generally speaking, with the decrease of sampling ratio, the performance of these algorithms has almost the same trend, but the ranking of these algorithms does not change. That is clear in the Karate network. In the other three small networks, though the performance of a few algorithms is similar (e.g., RA, AA, and CN in Dolphins and Word, CN, and LP in Metabolic), the stability of algorithm rankings is also remarkable. As for the other four large networks, because the neighborhood-based algorithms have very similar performance, we regard them as one class. The same result can be obtained from Figure 4 that the ranking of LP, PA, and the neighborhood-based algorithms is stable in a network across sampling ratios.

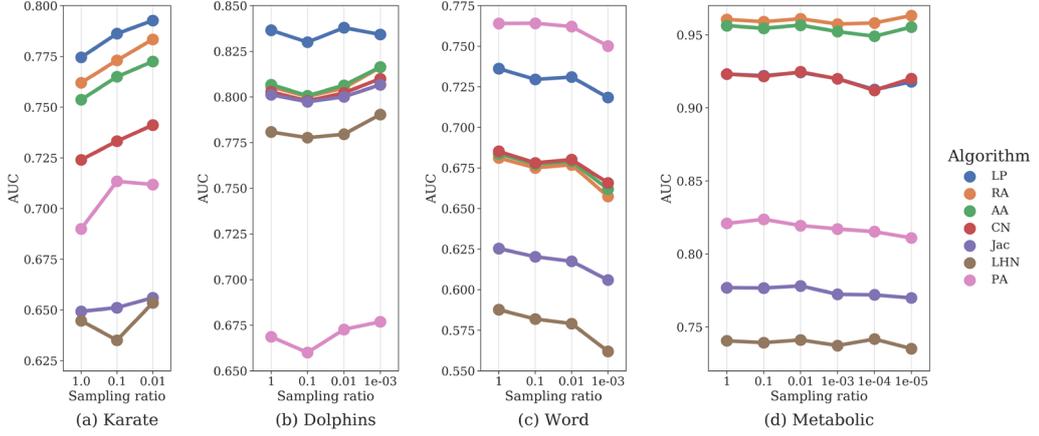

Figure 5. Rankings of link prediction algorithms in small networks.

The stability of algorithm ranking stems from two facets. First, the sampling of unobserved links does not change the performance of these algorithms significantly, and the AUC remains unchanged when the network size grows. Second, these neighborhood-based algorithms have similar performance due to the problem of "degeneracy of the states" mentioned above. Therefore, they show the same upward or downward trend when the AUC is affected by sampling. Empirically, the ranking results demonstrate that the sampling strategy is feasible in evaluating link prediction algorithms.

*5.4 Variance of AUC*

To further verify our theory and corollaries, we plot the variance of the AUC of seven link prediction algorithms in Figure 6. Here, we are concerned about how the performance variance changes with the sampling ratio in these networks. Hence, what is the most appropriate sampling ratio for a network?

There are two obvious tendencies. First, the performance variance of each algorithm in all networks goes up with the decline of the sampling ratio, which is consistent with Theory 1. Yet, the upward trend is divided into two phases by a turning point. Before the turning point (high sampling ratio), the performance variance changes little, and after the turning point (low sampling ratio), it rises sharply. The turning point is positively correlated with network size. For instance, the performance variance increases dramatically after the sampling ratio below 0.1 in Karate, Dolphins, and Word, whereas the cut-point in TVShows, LastFM, and Company is $10^{-5}$ (PA is $10^{-4}$). The results show that there is a lower limit for the feasible sampling ratio, which decreases with the increase of network size (in line with Corollary 2). Hence, we can empirically conclude that it is appropriate for any network when the sample number of unobserved links is not less than the number of links in the probe set.

Second, the performance variance of each algorithm falls significantly with the growth of network size, which is consistent with Corollary 3. For example, the performance variances of these algorithms are between $10^{-3}$ and $10^{-2}$ in Karate, Dolphins, and Word, and drop to about $10^{-5}$ in TVShows, LastFM, and Company. That implicates that the AUC of link prediction algorithms becomes more stable in large networks, which confirms our previous observations in Section 5.2.

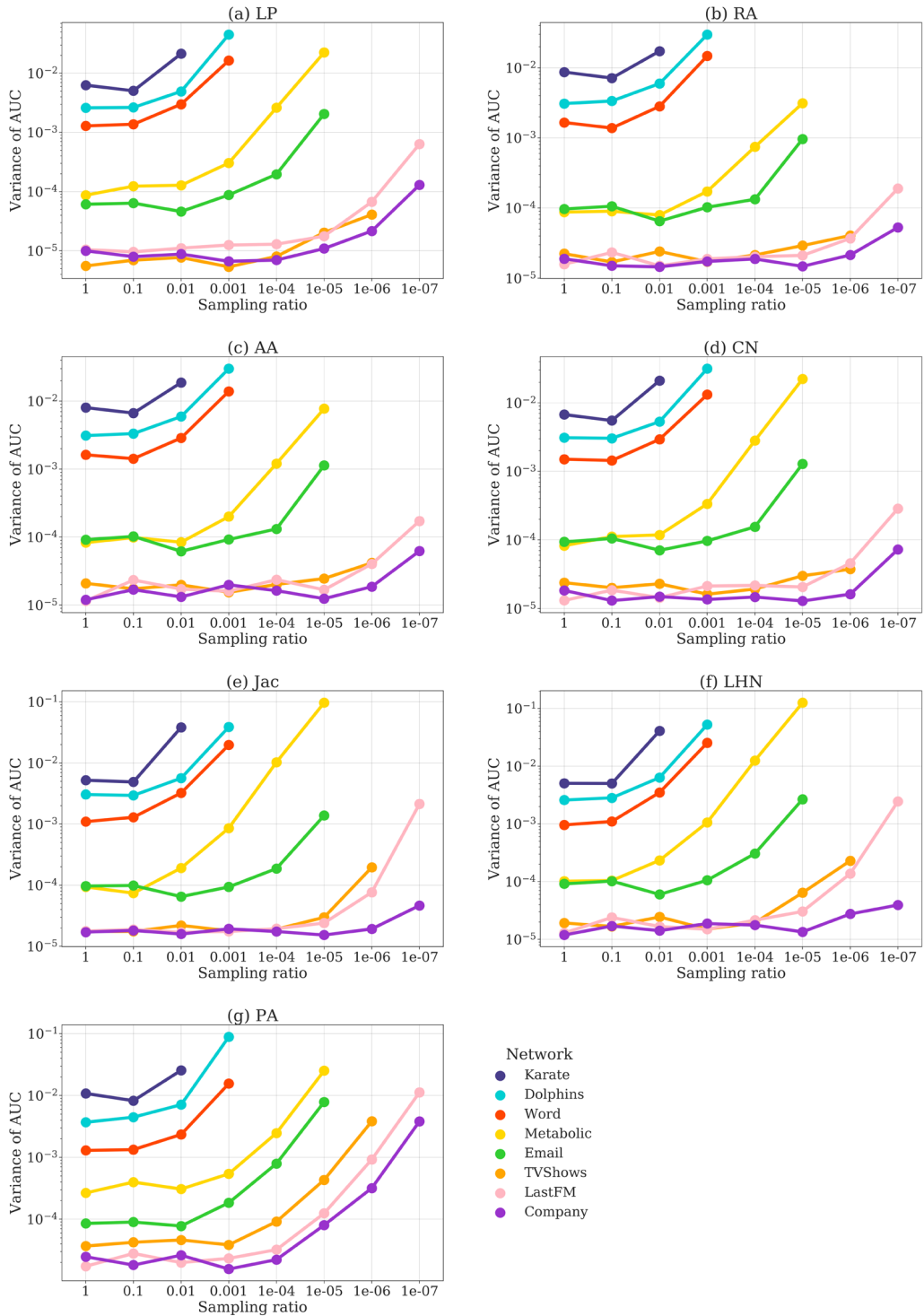

Figure 6. Performance variance of seven link prediction algorithms in eight networks.

Taken together, our results show that it is more reliable for the evaluation of unsupervised link prediction in large networks than that in small networks. Moreover, it is feasible to sample unobserved links in large networks with a wide range of sampling ratio.

## 5.5 Evaluation Time

Here, we focus on the effect of sampling on evaluation time, in other words, to what extent sampling unobserved links can reduce the evaluation time. Figure 7 shows the running time for evaluating link prediction algorithms in eight networks under different sampling ratios.

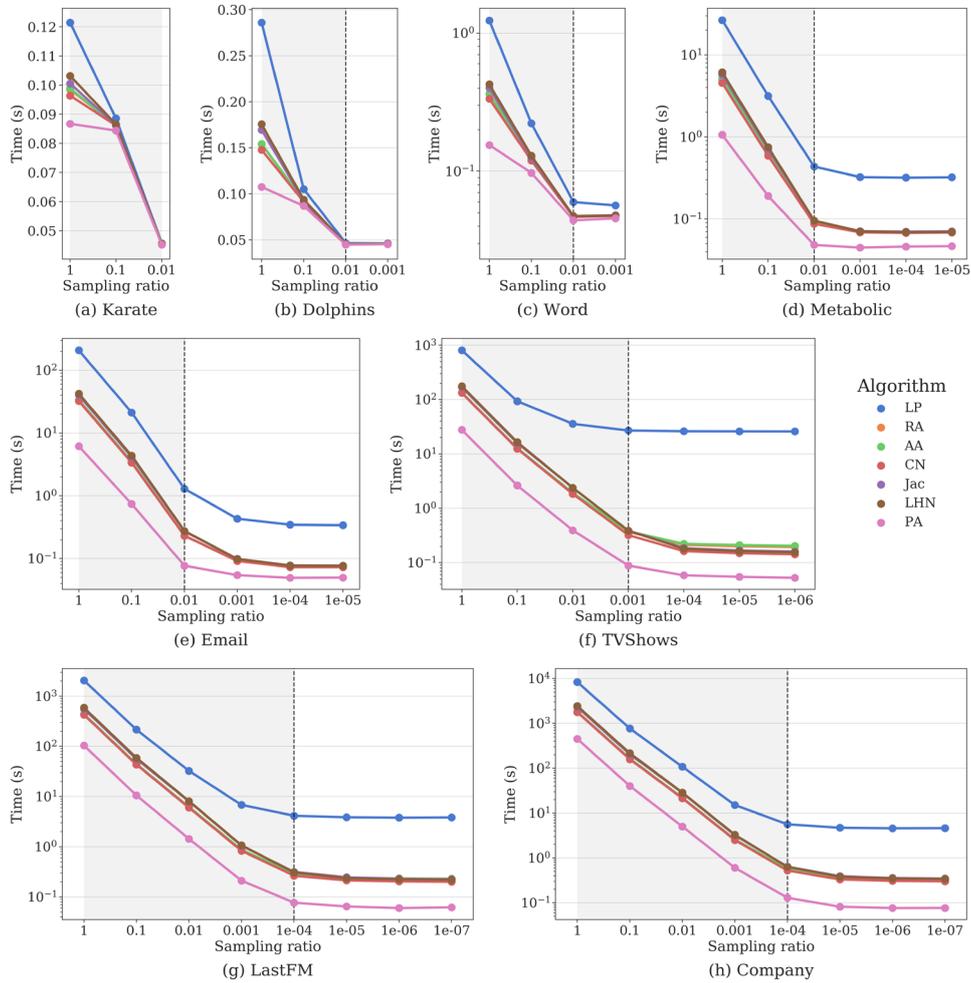

Figure 7. Evaluation time of seven link prediction algorithms in eight networks across sampling ratios.

Generally, different from the performance variance, the evaluation time of algorithms reduces with the decline of the sampling ratio. But similar to Figure 3, the downward trend also has two phases. At first, the evaluation time of an algorithm exhibits linear decline and then remains nearly constant. That occurs because when the sampling ratio is high, the evaluation time mainly depends on the number of sampled unobserved links that heavily outnumber probe links; when the sampling ratio becomes low, the evaluation time depends on the number of probe links that is a constant. Taking the Email network as an example, the number of sampled unobserved links is 635 when the sampling ratio is 0.001, which is close to the number of probe links, 545. When the sampling ratio is larger, the number of sampled unobserved links is more than the number of probe links so that the time increases. When the sampling ratio is lower than 0.001, the number of unobserved links sampled is almost negligible compared with the number of probe links. Thus, the evaluation time remains stable.

Not surprisingly, algorithms that use more structural information take longer. LP always takes the longest time since it requires the most information among these algorithms; whereas PA that requires the least information is the most time-saving algorithm in all networks.

*5.6 The balance between performance and time*

Based on the above findings, we can strike the right balance between performance and time of the evaluation for unsupervised link prediction. Specifically, for a network, there exists a sampling ratio to minimize the evaluation time while keeping the performance variance nearly constant. Figure 8 shows two examples of the LP algorithm.

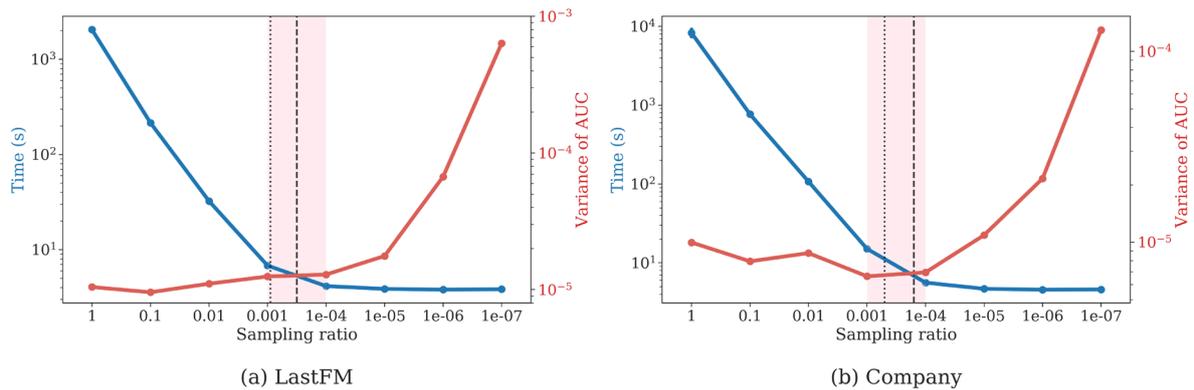

(a) LastFM  (b) Company

Figure 8. The performance variance and evaluation time of LP. (a) LastFM network and (b) Company network. Red lines: the variance of AUC; blue lines: the evaluation time; black dotted lines: $|N_s| = |E|$; black dashed lines: $|N_s| = |V|$.

In the LastFM network, when the sampling ratio is between $10^{-3}$ and $10^{-4}$, the performance variance of LP changes little, and the evaluation time of LP is reduced to 10 seconds, less than 1% of the time without sampling. That shows that there is a sampling ratio region (pink region), which makes the algorithm stable and takes less time. Moreover, when the number of unobserved links sampled is equal to the number of observed links (black dotted line) or the number of nodes (black dashed line), the corresponding sampling ratio, $|E|/|N|$ or $|V|/|N|$, is just in this region. That also occurs in the Company network. The results show that the time complexity of the evaluation paradigm can be reduced to $O(|E|)$ and even $O(|V|)$. Surprisingly, one can obtain the same performance by using an extremely small part (0.1% or 0.01%) of unobserved links.

## 6. Conclusion

In this paper, we asked one important question. Whether the evaluation conundrum in unsupervised link prediction can be addressed? Through theoretical analysis and extensive experiments, we found that the problem can be solved by sampling unobserved links. But the effect of sampling is somewhat different in small networks and large networks. In particular, it is risky and unnecessary in small networks (nodes<1000). First, our results show that the ranking of algorithms is stable in small networks across sampling ratios, but the performance variance of algorithms is significantly greater than that in large networks. Second, the time consumption of most link prediction algorithms is acceptable in small networks, and the requirement for computing resources is rare. However, it is feasible and inescapable to sample unobserved links in large networks. Our results indicate that the performance of algorithms is very stable even at a low sampling ratio, and the computational time degradation caused by this is remarkable. The sampling ratio should not be too small, and the recommended sampling ratio is about $|E|/|N|$ or $|V|/|N|$, that is, the number of unobserved links sampled is approximate to the number of observed links or nodes.

Our findings have broad implications for link prediction. First, many previous proposed algorithms only are evaluated in small networks. The results reported may be unreliable and misleading since algorithms exhibit large performance variance in small networks. Some algorithms (e.g., RA and AA) outperform other algorithms by a large margin in small networks, but they may be unable to maintain this advantage in larger ones. It is necessary to reevaluate their real performance in large networks. Second, for the sake of testing the proposed algorithm in large networks, previous studies have to sample a subnetwork from the original network to perform this evaluation due to the computational limitation. Yet, the topological properties in the subnetwork are quite different from those in the original network, leading to a serious loss of topological information. So this last resort is not a feasible solution. Future studies can use the sampling unobserved links strategy to evaluate proposed algorithms in large networks directly.

In sum, our study challenges the evaluation conundrum of unsupervised link prediction and shows that sampling unobserved links strategy is an effective and efficient method to solve this difficult problem. We hope that our findings will shed light on potential future studies for link prediction.

## CRediT authorship contribution statement

**Jingwei Wang:** Methodology, Software, Writing - original draft. **Yunlong Ma:** Formal analysis, Resources, Supervision, Funding acquisition. **Yun Yuan:** Validation, Software, Resources.

## Declaration of Competing Interest

The authors declare that they have no known competing financial interests or personal relationships that could have appeared to influence the work reported in this paper.

## Acknowledgments

This work was supported in part by the National Key R&D Program of China (Grant no. 2019YFB1704700), the National Natural Science Foundation of China (Grant nos. 61873191, 61973237, and 7207041147), and the Science and Technology Commission of Shanghai Municipality (Grant nos. 19JG0500700 and 20JG0500200).

## Data and Code availability

Data and code necessary to reproduce all results and plots will be made freely available.